\documentclass[doublecol]{epl2} 

\usepackage{dcolumn}
\usepackage{booktabs}
\usepackage{color}
\usepackage[all]{nowidow}
\usepackage{csquotes}
\usepackage{amsmath}
\usepackage{csquotes}
\usepackage{amsfonts}
\usepackage{mathtools}
\title{Particle production in $p$-O collisions at LHC energy}
\shorttitle{Particle production in p-O collisions} 

\author{Chuan Li \and Georg Wolschin}
\shortauthor{C. Li, G. Wolschin}

\institute{                   
Institute for Theoretical Physics, Heidelberg University, Philosophenweg 16, Heidelberg, 69120, Germany
}

\abstract
{We compare our previous predictions for charged-hadron production in $p$-O collisions at an energy of $\sqrt{s_\mathrm{NN}}=9.618$~TeV {as published in 2025}  with preliminary Run 3 data from ALICE and adapt the model parameters to the data.
			Our three-source model comprises a gluon-gluon central source and two {valence-quark soft-gluon} fragmentation sources. We describe the initial conditions using color-glass-condensate states and calculate the time-dependent partial thermalization with a relativistic diffusion model. 
			An inversion of the maximum production amplitude at larger pseudorapidities from backward to forward towards peripheral collisions is predicted.}

\begin{document}

\maketitle

\section{Introduction}
\label{intro}

Charged-hadron production in collisions of asymmetric systems such as d-Au at the BNL Relativistic Heavy Ion Collider (RHIC) \cite{alver11} or $p$-Pb at the CERN Large Hadron Collider (LHC) \cite{alice23} has successfully been accounted for in a three-source relativistic diffusion model. With a forward-going fragmentation source that accounts for {valence-quark soft-gluon interactions}, a central gluon-gluon source, and a backward-going fragmentation source, we had reproduced RHIC data for pseudorapidity distributions of charged hadrons in $\sqrt{s_\mathrm{NN}}=200$ GeV d-Au collisions and their detailed centrality dependence \cite{biya05}. 

At that time, the initial conditions were assumed to be $\delta$-functions in rapidity space. With physically more realistic color-glass-condensate (CGC) initial conditions, we have recently described charged-hadron pseudorapidity distributions at various centralities in 5.02 TeV and 8.16 TeV $p$-Pb collisions, and determined the model parameters \cite{sgw24} in {$\chi^2$} minimizations with respect to ATLAS \cite{atlas16} and ALICE \cite{adam15,alice24} data. 

As part of a dedicated small-system run at LHC in 2025, the astrophysically relevant $p$-O system has been investigated at $\sqrt{s_{NN}}=9.618$ TeV. Within our three-source model, we predicted centrality-dependent charged-hadron pseudorapidity distributions \cite{ksw25} before data became available.
With the overall magnitude for our prediction taken from minimum-bias Monte Carlo calculations \cite{piwe23}, we 
calculated the centrality-dependent distribution functions in pseudorapidity space. Since preliminary $p$-O data have meanwhile become available \cite{uri25}, we compare in this Letter our centrality-dependent predictions with these data, readjust the model parameters, and sharpen our prediction at large pseudorapidites where no data are available yet.

In our three-source model as applied to $p$-O, colour-glass states provide the initial conditions for the two fragmentation sources based on hybrid factorization. The subsequent partial thermalization of the fragmentation distribution functions in rapidity space is accounted for 
 in a relativistic diffusion model.  The central-rapidity source is based on $k_{T}$-factorization. Results of pseudorapidity distributions for produced charged hadrons in $\sqrt{s_{NN}}=9.618$ TeV
  $p$-O collisions are calculated in eight centrality classes -- as in the ALICE experiment -- in the full pseudorapidity range and compared with ALICE preliminary data, as well as with our previous predictions. 
  \section{Three-source model}
\label{initial}
We construct the distribution of charged-hadron production in rapidity space as an incoherent superposition of three sources \cite{biya05,sgw24} 
\begin{align}
  &  \frac{dN_\text{ch}}{dy}(y,\tau_\text{int}) = N_\text{ch}^1 R_1(y, \tau_\text{int}) + N_\text{ch}^2 R_2(y, \tau_\text{int})\\\nonumber
& \hspace{5.2em} + N_\text{ch}^{gg} R_{gg}(y)\,.
\end{align}
{As stated in \cite{biya05} for d-Au at $\sqrt{s_\mathrm{NN}}=200$~GeV, the distribution function $R_1(y, t)$ accounts for charged hadrons produced in the forward (here: proton-going) source, whereas $R_2(y, t)$ is the backward (here: oxygen-going) source. In the original model, we had used $\delta$-function initial conditions  that allowed us to provide analytical solutions. With physically more realistic color-glass initial conditions as in \cite{sgw24} for $p$-Pb at LHC energies that we also use in the present work for $p$-O, the time evolution of the fragmentation sources can only be modeled numerically.

The fragmentation sources} undergo partial thermalization {through particle creations and collisions. They} are integrated up to the interaction (freezeout) time $\tau_\text{int}$. The number of charged hadrons in the fragmentation sources are  $N_\text{ch}^{1,2}$ {-- these are free parameters in the model, which will be determined in {$\chi^2$ or least-square} minimizations with respect to data if available, or extrapolated from other collision systems such as $p$-Pb where data have already been taken.} $N_\text{ch}^{gg}$ corresponds to the production of charged hadrons in the central gluon-gluon source {that has a distribution function $R_{gg}(y)$ in rapidity space. As in \cite{sgw24}, we model it as a color-glass distribution arising from gluon-gluon collisions.}

The time evolution of the fragmentation sources towards partial equilibration  in $p$-O collisions at $\sqrt{s_\text{NN}}=9.618$ TeV has been investigated in \cite{ksw25}. It is based on the numerical solution of a Fokker--Planck equation
\cite{HIC,hgw24,extrapol} for the distribution functions $R_{1,2}(y,t)$ in rapidity space
\begin{align}
    \label{FPE}
 &   \frac{\partial}{\partial t}R_{1,2}(y,t) =\\\nonumber
  &   - \frac{\partial}{\partial y} [J_{1,2}(y)R_{1,2}(y,t)] + D^{1,2}_y \frac{\partial^2}{\partial y^2}R_{1,2}(y,t)\,.
\end{align}
The rapidity $y= \ln \left[ {(E+p)/(E-p})\right]$ is determined by the particle's energy $E$ and its momentum $p$.
{We define the $p$-going beam rapidity as positive and the O-going direction as negative, and we have adapted} the preliminary ALICE $p$-O data \cite{uri25} accordingly.

The rapidity diffusion coefficients $D^{1,2}_y$ cause the broadening of the distributions that arises from microscopic physics. We assume $D^{1,2}_y$ to be constant over time and rapidity. The drift terms $J_{1,2}(y)$  are  responsible for the shift of both fragmentation sources in rapidity space  towards the equilibrium value. To obtain a Maxwell-Jüttner distribution as the stationary solution of Eq.\,(\ref{FPE}), drift terms of the form $ J(y) = - (m_T D_y/T)\sinh (y)$ would be required \cite{HIC}.
The prefactor in this fluctuation-dissipation relation depends on the diffusion coefficient $D_y$, the transverse mass $m_T$, and the equilibrium temperature  $T$. 

Here, we use a linear drift term  \cite{biya05,sgw24} resulting as the leading-order approximation, $J(y) =(y_\text{eq}-y)/\tau_y$,
with the equilibrium rapidity $y_\text{eq}$, and the rapidity relaxation times $\tau_y\equiv\tau^{1,2}_y$ governing the time scales over which the fragmentation distributions approach equilibrium. The Fokker--Planck equation with constant diffusion and linear drift corresponds to the so-called Uhlenbeck-Ornstein process \cite{UO}. The rapidity $y_\text{eq}$ at which equilibrium is reached is zero in symmetric systems, but non-zero in asymmetric systems such as the $p$-O collision. For sufficiently high beam rapidities, given at LHC energies, it can be calculated as  \cite{HIC}
\begin{align}
    y_\text{eq}(b)= \frac{1}{2} \ln \left( \frac{ \langle m_T^{(2)}(b) \rangle}{ \langle m_T^{(1)}(b)\rangle} \right)
\end{align}
with the centrality-dependent average transverse masses $\langle m_T^{(1,2)}(b) \rangle=[{m_p^2+\langle p_T^{(1,2)}(b)\rangle^2}]^{1/2}$. In equilibrium, the charged-hadron distribution approaches a Maxwell-Jüttner distribution in three-dimensional momentum space $(y,p_T)$ 
\begin{align}
    &E \frac{d^3N}{dp^3} \sim E \exp(-E/T) =\\\nonumber
    & m_T \, \cosh(y) \exp(-m_T \cosh(y)/T)\,,
\end{align}
and by
 integrating over the transverse mass $m_T$, one obtains the equilibrium limit for particle production in the longitudinal fragmentation sources \cite{sgw24}. 
Accordingly, the time-dependent distribution functions $R(y,t)\equiv R_{1,2}(y,t)$ that solve the Fokker--Planck equation are integrated over transverse mass to obtain the rapidity distributions $dN/dy(y,t)$. Here, we assume {cylindrical} symmetry in phase space for transverse momentum and transverse mass, to obtain 
\begin{align}
    E \frac{dN}{dy}(y,t) = c \int_m^\infty m_T^2 \cosh(y) R(y,t) \, dm_T\,,
\end{align}
where the constant $c$ absorbs all other constants appearing in the expression.

Although the Fokker--Planck equation with linear drift and constant diffusion can be solved analytically for $\delta$-function or Gaussian initial conditions, a numerical solution is required for more sophisticated initial conditions such as color-glass-condensate initial states that we use
 here, as in \cite{sgw24,ksw25}. It is carried out for the two fragmentation sources with a  \texttt{C++} code based on the finite-element method, originally written for the \mbox{$p$-Pb} system \cite{sgw24} and later modified for the $p$-O prediction \cite{ksw25}. We obtain the distribution functions for the two fragmentation sources, $R_{1,2}(y,t)$, and use the color-glass result for the central gluon-gluon source, $R_{gg}(y)$ \cite{ksw25}. We then construct the distribution of charged-hadron production as an incoherent superposition of the three sources due to the linearity of the Fokker--Planck equation.
\subsection{Initial conditions}
There are two distinct fragmentation sources for hadron production from interactions between valence quarks and soft gluons.  For the $p$-going source, these correspond to interactions between the valence quarks of the proton and the gluons of the oxygen ion, and vice versa for the O-going source. For the O-going side, we scale the valence-quark distribution function $f_{q/A}$ with the centrality-dependent number of participating nucleons $N_\text{part}$, which we determine with a Glauber code \cite{GlauberGIT,GITGlauberpaper}.

For the fragmentation sources, the initial distributions depending on rapidity $y$ and transverse momentum $p_T$ are determined using the method of hybrid factorization \cite{hybrid}. 
The cross section includes the valence-quark distribution function and the gluon distribution for the separately treated incoming nuclei \cite{hybrid}.

The initial state for single-inclusive hadron production in asymmetric proton-nucleus scattering becomes \cite{dhj06,alt11}
\begin{align}
    &\frac{d^3N_{qg}^h}{dy \, dp_T^2} = \frac{1}{(2 \pi m_T)^2}\\\nonumber
    &\times \int_{x_F}^1 \frac{dz}{{z^2}} D_{h/q}(z,\mu^2_f) x_1 f_{q/p}(x_1,Q_f^2) \varphi(x_2,q_T^2)
    \label{inistopp}
\end{align}
with Bjorken $x$ of the valence quark $x_1$ and that of the soft gluon $x_2$. The produced hadrons are $h= \pi, K, 
p$, and their antiparticles. The factorization scale is $Q_f^2=p_T^2$. Feynman $x_F$ is defined by the transverse momentum of the parton $k_T$ and that of the produced charged hadron $p_T$,
$ x_F = xp_T/k_T$.
The fraction of quark energy carried by the produced hadron is
   $ z(x) = x_F/x$,
with the boundary conditions $z(1)=x_F$ and $z(x_F)=1$, and the associated differential $dz$,
 $   dx/x_F= - (x^2/x_F^2) dz = - dz/z^2$.
The gluon distribution function depends on the effective transverse momentum 
$  q_T = m_T/z=\sqrt{(p_T^2+m_p^2)}/z$.
The fragmentation function $D_{h/q}(z, \mu_F^2=p_T^2)$ gives the probability that a parton fragments into a hadron carrying the fraction $z$ of the quark's energy{, and we use the updated AKK08 results \cite{akk08}, with parameters provided separately for each hadron species, and symmetrical production of both hadrons and antihadrons}. 

We use the MSTW parametrization for the valence-quark distribution function $f_{q/A}(x_1,Q_f^2)$ \cite{mstw09},  and the KLN model \cite{Kharzeev2001b,Kharzeev2001, Kharzeev2004b,Kharzeev2005a,Duraes2016} for the gluon distribution function $\varphi(x_2,q_T^2)$ 
\begin{equation}\label{KLN_UGD}
\varphi_\mathrm{KLN}\left(x,k^2\right) :=
\begin{dcases}
\frac{2C_F}{3\pi^2\alpha_s(Q^2)},  & k^2\leq Q^2_s(x)  \\[8pt]
\frac{2C_F}{3\pi^2\alpha_s(Q^2)}\frac{Q_s^2(x)}{k^2}, & k^2>Q^2_s(x)\,,
\end{dcases}
\end{equation}
where $k^2$ defines the internal momentum transfer scale.
The gluon distribution functions are modified  according to \cite{gdfcorr}
as $ \hat{\varphi}(x,k^2) = (1-x)^4 \varphi(x,k^2)$
to avoid unphysical contributions from large values of $x$.
The gluon saturation scale is 
\begin{equation}
    Q_s^2(x) = A^{1/3} Q_0^2 \left( \frac{x_0}{x} \right)^{\lambda}\,.
    \label{saturationscale}
\end{equation}
For the $p$-going source, we set $A = 1$ and $x_0 = 1$. According to HERA data, the parameters for the proton are $\lambda = 0.288$ and $Q_0^2x_0^{\lambda} = 0.097 \,$GeV$^2$. These values were obtained from fits to the experimental results of deep-inelastic electron-proton scattering \cite{Q0scale}. In \cite{hgw24}, we have demonstrated the consistency of this parametrization with charged-hadron production at LHC energies for Pb-Pb collisions. Hence, we consider  these values to be applicable in $p$-O collisions as well. 

For the O-going source, we use $A=16$ and $x_0 = 1$, and the same value $\lambda = 0.288$ for the saturation-scale exponent. In line with our previous works \cite{sgw24,ksw25}, we account for the centrality dependence with an impact-parameter-dependent scale parameter $Q_0^2(b)$. For $p$-Pb, we had determined the centrality dependence at $\sqrt{s_{NN}}=5.02$ and 8.16 TeV rather precisely \cite{sgw24} in comparison with ATLAS and ALICE data, and we use these results as a guidance for the centrality dependence in $p$-O at 9.62 TeV.


{For the forward/backward initial particle production, we take into account only the distribution of incoming quarks on the large-$x$ side, thus neglecting the small contributions of incoming gluons and {sea quarks}, as we did earlier \cite{mtw09} when calculating the Pb-Pb net-proton (``stopping") distributions in comparison with SPS data.}  The initial CGC distributions that we use as initial conditions for the fragmentation sources are indeed equivalent to the stopping distributions, which exhibit already a sizeable shift towards mid-rapidity, as well as a substantial broadening, cf. Fig.\,1  in \cite{ksw25}. The final charged-hadron fragmentation distributions emerge in the course of partial equilibration as accounted for by the Fokker--Planck equation. This requires the transport parameters $D_y, J(y)$ and the interaction times $\tau_\text{int}$. Since the timescales are not observable, we make use of the full widths at half maximum of the fragmentation distributions, which are related to $D_y\tau_y$ and 
$\tau_\text{int}/\tau_y$, as well as the shifts of the maxima towards equilibrium, which depend on $\tau_\text{int}/\tau_y$.
\begin{center} 
\begin{table*}[h] 
\centering
\caption{Calculated impact parameters $b$ and mean number of participants $\langle N_\text{part} \rangle$ for the centrality classes in the ALICE $p$-O experiment at 9.62 TeV \cite{uri25}. The centrality-dependent parameters $Q_0^2$ determine the gluon saturation scale.\\}
{
\begin{tabular}{c|ccccccccc} 
\hline
cent.[\%] & 0-5 & 5-10 & 10-20 & 20-30 & 30-40 & 40-50 & 50-60 & 60-90 \\ 
\hline
$b \,$[fm] & 0-0.9 & 0.9-1.3 & 1.3-1.9 & 1.9-2.3 & 2.3-2.7 & 2.7-3.0 & 3.0-3.3 & 3.3-4.0 \\
$\langle N_\text{part} \rangle$ & 6.9& 6.2& 5.2& 4.2& 3.5& 3.0 & 2.7 & 2.3\\ 
$Q_0^2$ [GeV$^2$] & 0.031 & 0.029& 0.026 & 0.022 & 0.018 & 0.015 & 0.013 & 0.010\\
\hline
\end{tabular}
}
\label{tab1}
\end{table*}
\end{center}

The centrality-dependent numbers of participants are obtained using a Glauber code \cite{GlauberGIT} that is based on \cite{GITGlauberpaper}. 
For the inelastic nucleon-nucleon cross section, we use $\sigma_{NN} = 75\,$mb for 9.62 TeV $p$-O. This value is determined from LHCb data for the inelastic proton-proton cross-section \cite{ppcrosssec}, in accordance with measurements by  ALICE, ATLAS, and TOTEM. 
We take an exponential profile for the charge density distribution of the proton,
$ f(r) = \exp \left( -r/R \right)$ 
with half-density radius $R=0.234 \,$fm \cite{GlauberGIT}. 
For the oxygen nucleus, we use a three-parameter Eerhart-Fermi distribution \cite{GITGlauberpaper} 
\begin{align}
    f(r) = \frac{1 + W \frac{r^2}{R^2}}{1 + \exp(\frac{r-R}{a})} 
\end{align}
with $R=2.608 \,$fm for the half-density radius, $a=0.513 \,$fm for the skin parameter, and $W=-0.051$ for the suppression near $r=0$. 

\begin{figure}[!h] 
\centering 
\includegraphics[width=\columnwidth]{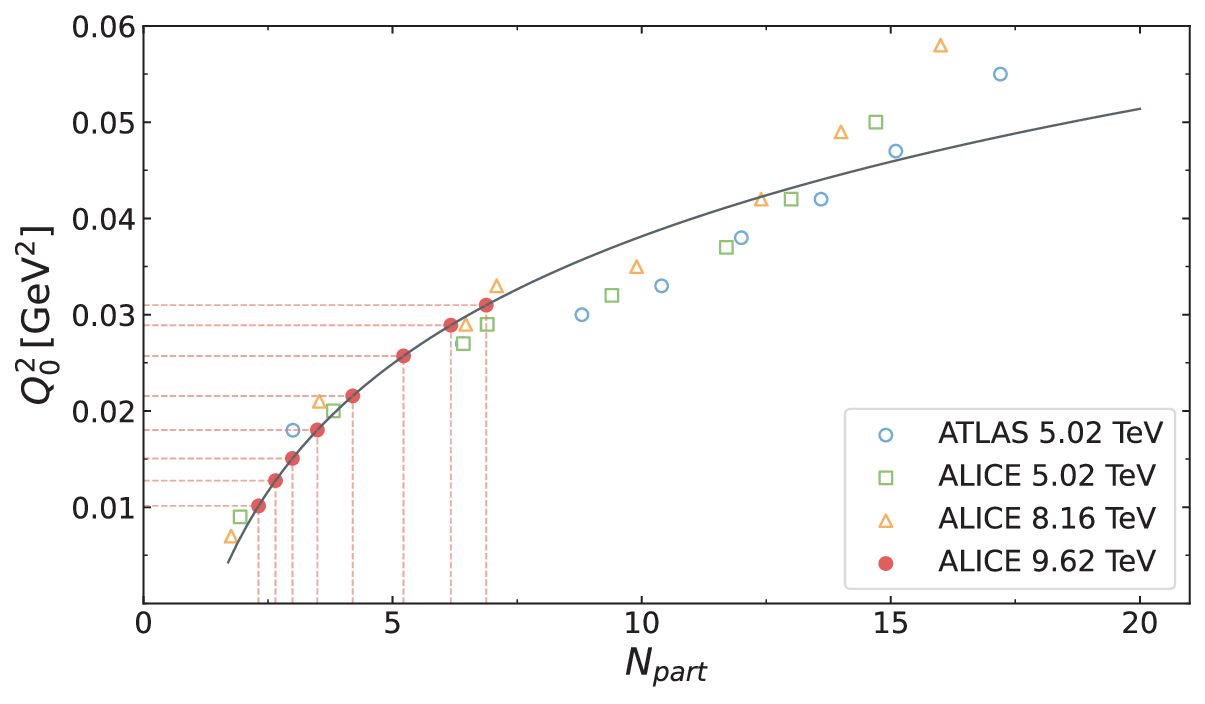} 
\caption{Centrality dependence of the saturation-scale parameter $Q_0^2$
with values determined for $p$-Pb collisions at incident energies $\sqrt{s_{NN}}=5.02$ TeV, 8.16 TeV from
\cite{sgw24}. The solid line serves as a two-parameter fit to all data sets, see text. The solid points mark the values
we take for $p$-O collisions in the eight new centrality bins of the ALICE experiment \cite{uri25}.} 
\label{fig1}
\end{figure}

The resulting mean numbers of participants $\langle N_\text{part} \rangle = 1 + \langle N_\text{part}^\text{O}\rangle$ are shown in table\ref{tab1}  for eight centrality classes as in the ALICE experiment \cite{uri25} with the corresponding impact parameters $b$, updated from our earlier prediction \cite{ksw25}, which was in seven slightly different centrality classes as in the $p$-Pb experiment. 
The functional form for the centrality dependence of the parameter $Q_0^2$ as obtained from a {$\chi^2$} minimization \cite{sgw24} with respect to $p$-Pb data  is expressed {in form of the empirical relation} 
\begin{align}
\label{eq7}
    Q_0^2 = c \cdot \log(N_\text{part}) + d
\end{align}
with $c= 0.019 \,$GeV$^2$ and $d=-0.006 \,$GeV$^2$. As shown in Fig.\;\ref{fig1}, this indicates a significant decrease of $Q_0^2$ for small values of $N_\text{part}$, especially in the relevant range for $p$-O collisions. 
Table\,\ref{tab1} includes the centrality-dependent parameters $Q_0^2 (\langle N_\text{part} \rangle)$ that determine the local gluon saturation scale according to 
Eq.\, (\ref{saturationscale}) as $Q_\text{s}^2(b)=Q_0^2(b)\times 16^{1/3}/x^\lambda$  with $\lambda = 0.288$ from HERA. {The corresponding gluon saturation momentum is a model-specific parameter, and care should  be taken when comparing results based on this definition with those obtained using other conventions}.

\subsection{Central source}
For the central gluon-gluon source we use $k_T$ factorization, as is common for charged-hadron production at high center-of-mass energies $\sqrt{s}$ corresponding to low Bjorken $x$ of the partons. The inclusive cross section was originally presented in \cite{initialgg}. It has been modified in \cite{sgw24,ksw25} for asymmetric $p$-Pb and $p$-O collisions by introducing two distinct {additive} subprocesses that each obey $k_T$ factorization and are proportional to the respective numbers of participants $N_1, N_2$. The initial state for the central gluon-gluon source with respect to rapidity $y$ and transverse momentum $p_T$ of the produced hadron {becomes}
\begin{align}
&\frac{d^3N_{gg}^h}{dy \, dp_T^2}= \frac{2\alpha_s}{C_F \, m_T^2}\\ \nonumber
&     \times \int_0^{p_T}dk_T^2 \big[ \,
        N_1 \varphi_1(x_1,k_T^2) \varphi_2(x_2, |\mathbf{p}_T-\mathbf{k}_T|^2) \notag \\ \nonumber
 &    + N_2 \varphi_1(x_2,k_T^2) \varphi_2(x_1, |\mathbf{p}_T-\mathbf{k}_T|^2) \big] \text{.}
\end{align}
We calculate the centrality dependence of $N_2$ as in case of the fragmentation sources,  using a Glauber model for each centrality class. The Casimir factor is $C_F = (N_C^2 - 1) / (2N_C) = \frac{4}{3}$ and the transverse mass {$m_T=(m_p^2+p_T^2)^{1/2}$}. The running of the strong coupling $\alpha_s$ is parametrized \cite{alpha_s} as
\begin{equation}
    \alpha_s(k^2) = \frac{4 \pi}{\beta \ln{(4k^2/\Lambda_\text{QCD}^2 + \mu})}
\end{equation}
with $\beta = 11 - \frac{2}{3} N_f = 9$, for the number of quark flavors $N_f =3$. The QCD-scale parameter is $\Lambda_\text{QCD} =0.241 \,$GeV and $\mu=16.322$. The latter follows from the boundary condition $\alpha_s(0) = 0.5$
and regulates the strong coupling at large dipole sizes $r\rightarrow \infty$ ($k\rightarrow 0$). 

The distribution depends on the unintegrated gluon distribution functions $\varphi_{1,2}(x, k_T^2)$ for the gluons in the proton $\varphi_1$ and in oxygen $\varphi_2$ with the transverse momentum of the gluon $k_T^2$ and  $x_1 = (m_T/\sqrt{s_{NN}})\,e^y, x_2 = (m_T/\sqrt{s_{NN}})\,e^{-y} $. Again we use Eq.\,(\ref{KLN_UGD}) for the gluon distributions.
\begin{figure}[!h] 
\center
	\includegraphics[width=\columnwidth]{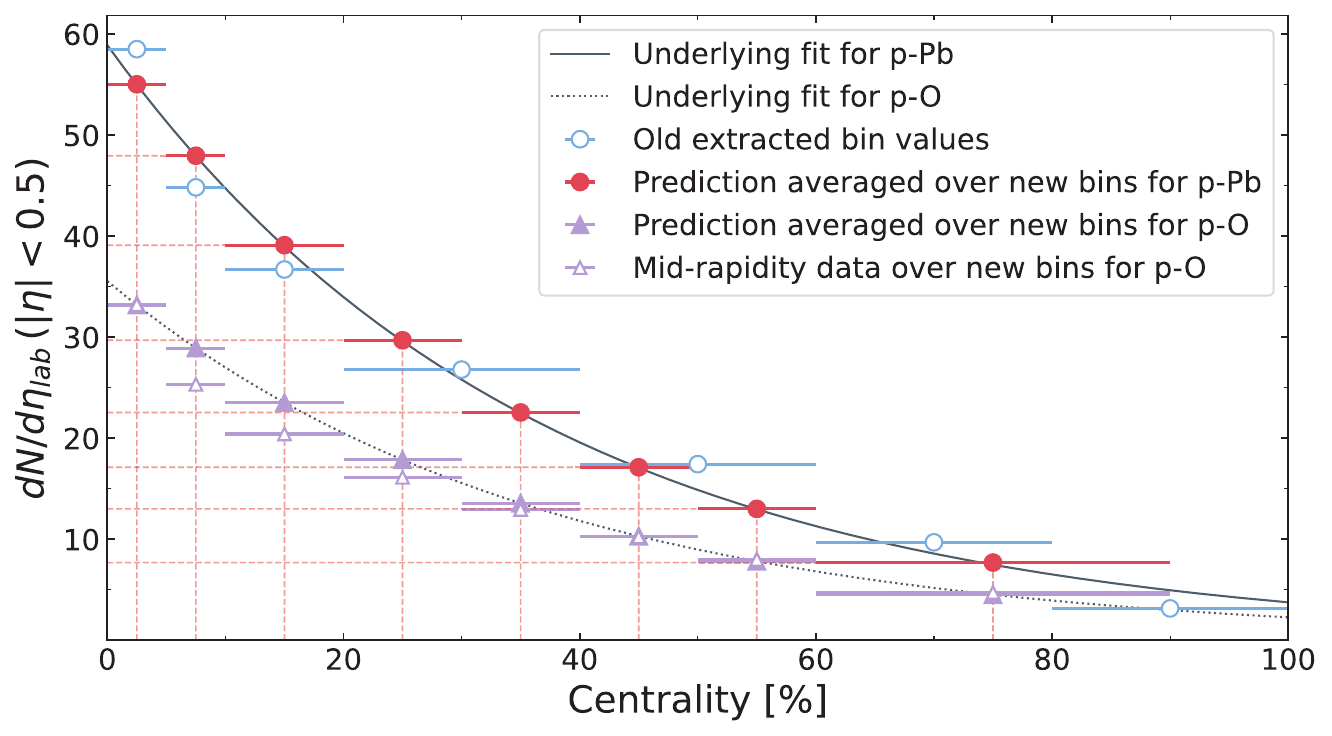}%
	\caption{\label{fig2}%
Interpolation of the mid-rapidity $dN/{d\eta}$ values for produced charged hadrons in $p$-Pb collisions at $\sqrt{s_{NN}}$=8.16 TeV according to \cite{sgw24} (solid curve),
 and the present $p$-O results at 9.62 TeV (dotted curve). The amplitude for $p$-O is set to the mid-rapidity data in the $0-5\%$
class, results together with preliminary ALICE mid-rapidity data \cite{uri25} are shown in all eight centrality classes. }
\end{figure}

\subsection{Superposition of the sources}
Especially for asymmetric systems, the superposition of the sources is very sensitive to the values of the transport parameters, {which depend on the system size $(\propto A^{1/3})$ and the available energy. The latter is rather similar in $p$-Pb $(\sqrt{s_{NN}}$=8.16 TeV) and $p$-O $(\sqrt{s_{NN}}$=9.62 TeV).} 
To estimate the transport parameters for $p$-O collision at $\sqrt{s_{NN}} = 9.618 \,${TeV}, their values in $p$-Pb at $\sqrt{s_{NN}} = 8.16 \,${TeV}, resulting from $\chi^2$ 
fits to ALICE data, are taken from \cite{sgw24}, and scaled in each centrality class 
to account for the different system sizes and time scales \cite{ksw25}.
\begin{figure*}[!ht] 
\centering
	\includegraphics[width=1.83\columnwidth]{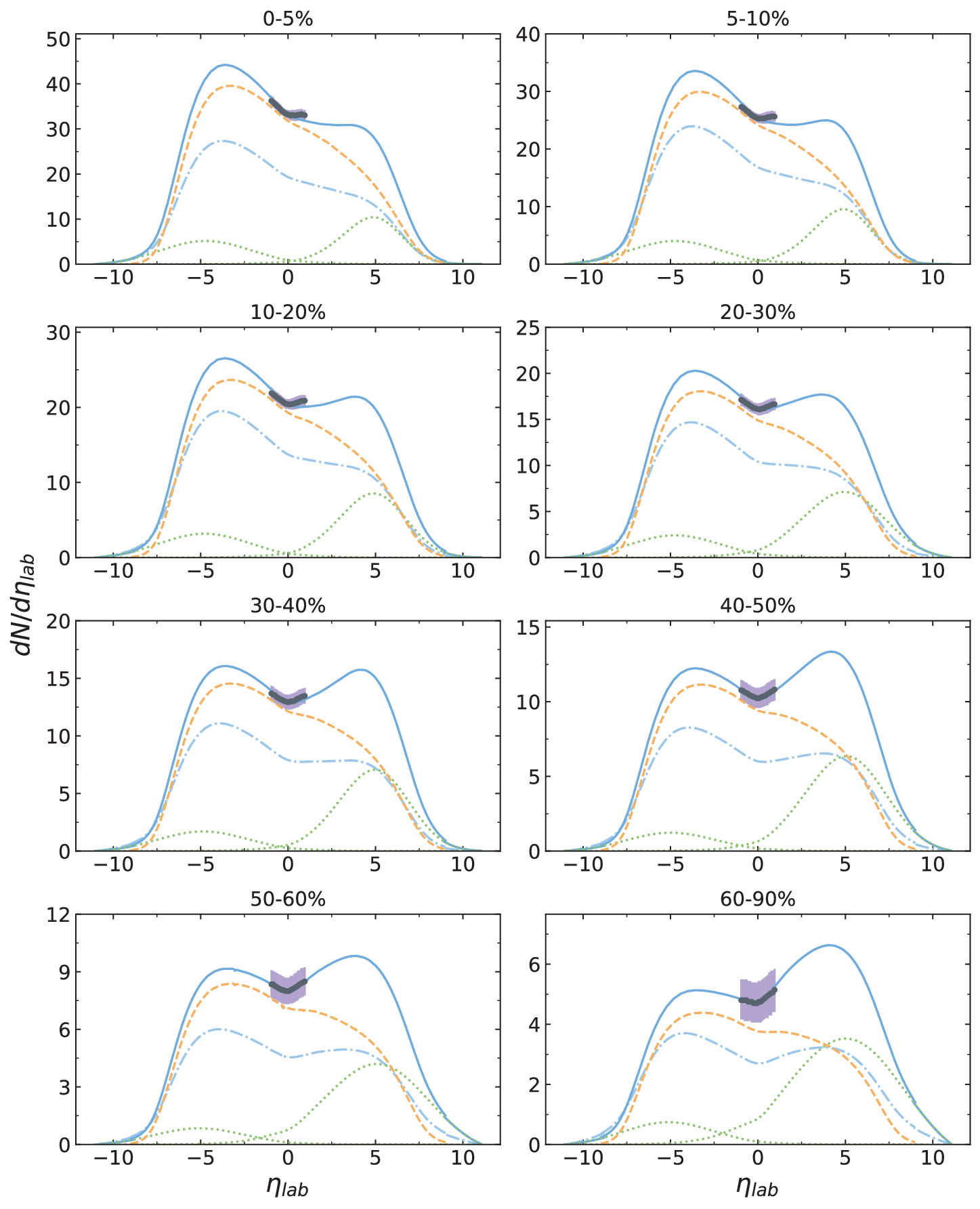}%
	\caption{\label{fig3}%
Calculated pseudorapidity distributions of produced charged hadrons (solid {blue} curves) in $p$-O collisions at $\sqrt{s_{NN}}=9.618$ TeV for eight centrality classes. The dashed {orange} curves show charged hadrons (pions, kaons, protons, and their antiparticles) from the central gluon-gluon source, the dotted {green} curves are the ones from the respective $p$-going (forward) and O-going (backward) fragmentation sources. The preliminary ALICE data are from \cite{uri25}, {with the shaded areas representing systematic uncertainties}. The dash-dotted {blue} curves are the distributions from our previous
prediction \cite{ksw25}, whereas the overall {amplitudes are} now adjusted to the preliminary {mid-rapidity} data. Note the inversion of the maximum particle-production amplitude from backward to forward towards more peripheral collisions.
}
\end{figure*}
\begin{table*}[h] 
\centering
\caption{Centrality-dependent parameters for charged-hadron production in 9.62 TeV $p$-O collisions:
transport parameter for the $p$-going source, 
$D^p \tau_y$, ratio of hadrons produced in the central source to those produced in the fragmentation sources, $R_{qg}^{gg}$, and ratio of particles originating from the O-going fragmentation source to those from the $p$-going source, $R_{p}^\text{O}$. $N_\text{ch}$ is  the total number of charged hadrons  produced by all sources in $p$-O collisions at each centrality.\\}
\begin{small}
{
\begin{tabular}{c|ccccccccc} 
\hline
centrality [\%] & 0-5 & 5-10 & 10-20 & 20-30 & 30-40 & 40-50 & 50-60 &60-90 \\ 
 \hline
$D^p \tau_y$ & 3.6 & 3.6 & 5.7 & 6.4 & 8.0 & 9.6 & 14.0 &19.5 \\ 
$R_{qg}^{gg}$ & 4.98& 4.42& 3.87& 3.32& 3.32& 2.76& 2.76 & 1.66\\
$R_{p}^\text{O}$ & 0.71& 0.61&0.48& 0.37 & 0.30& 0.21 & 0.18 & 0.18  \\
$N_\text{ch}$ & 468 & 365 & 302 & 245 & 203 & 167 & 128 & 81 \\  
\hline
 \end{tabular}
 }
\end{small}
\label{tab2}
\end{table*}
\section{Pseudorapidity Distributions}
To compare our theoretical results to data, a Jacobian transformation to pseudorapidity space must  be performed \cite{biya05,sgw24}, and the rapidity shift from the nucleon-nucleon center-of-mass frame to the laboratory frame has to be considered. In the ultrarelativistic limit that is applicable at TeV energies, the shift is given by
\begin{align}
   \Delta y=\frac{1}{2}\ln\left[\frac{A_1 Z_2}{A_2 Z_1}\right] \,,
    \label{shift}
\end{align}
yielding $\Delta y=0.347$ for $p$-O, irrespective of energy.
The beam rapidities are $|y_\text{beam}^p| = 9.582$ for the proton and $|y_\text{beam}^\text{O}| = 8.889$ for oxygen. The beam rapidity in the nucleon-nucleon frame of reference can be obtained by using $|\mathbf{p}_{NN}| =\sqrt{s_{NN}}/2$ as momentum per nucleon of the colliding beams, resulting in $|y_\text{beam}| = 9.235$. 

By integrating the Fokker--Planck equation for the color-glass initial conditions within the relativistic diffusion model, charged-hadron distributions in rapidity space are obtained for the fragmentation sources. For the central gluon-gluon source, the color-glass initial states are sufficient to account for the charged-hadron distributions in rapidity space. 
The shape and position  of each source in pseudorapidity are then fixed. However, they must be scaled to obtain the total yield of charged hadrons and the contribution of the three sources to the total distribution in each centrality class.
To account for the correct contribution of each source, two parameters are considered \cite{sgw24}: the ratio of hadrons produced in the central source to the those produced in the fragmentation sources, $R_{qg}^{gg} = N_{gg} / N_{qg}$, and the ratio of particles originating from the O-going fragmentation source to those from the $p$-going source, $R_{p}^{O} = N_{qg}^\text{O} / N_{qg}^p$.
The latter ratio is obtained by comparing the integrals of the two fragmentation sources over the full pseudorapidity range. It is given by the initial conditions, and depends sensitively on the result of the Glauber calculation -- which may be unreliable for a small system.  The ratio may change throughout the partial thermalization process, and we use it as a fit parameter, cf. table\,\ref{tab2} -- with values that are {a significant factor of  0.2-0.3} smaller than what is obtained from the Glauber code \cite{GlauberGIT}.

The contribution of the central source in $p$-O collisions is expected to be {relatively} more significant than that in $p$-Pb.
The scaling behavior of the ratio $R_{qg}^{gg}$  can be understood in terms of cold nuclear matter effects, which reduce the relative contribution of the central $gg$ source as compared to the $qg$ and $gq$ fragmentation sources. Such effects depend on the size of the transverse overlap. To quantify the influence of the transverse area, we have used a scaling factor proportional to $A^{-2/3}$. 
{As shown in table\,\ref{tab2} in comparison with table\,V for 8.16 TeV $p$-Pb in \cite{sgw24}, the ratio of $R_{qg}^{gg}$ values for central collisions in $p$-O to $p$-Pb is $\simeq 5.5$, corresponding to the inverse ratio of the transverse areas.}  

Since the overall particle production amplitude cannot be determined a priori in our model, we had normalized it \cite{ksw25} to match the maximum of the results of corresponding QGSJETII-04 and SIBYLL 2.3d Monte Carlo simulations at 10 TeV $p$-O in the mid-rapidity region \cite{piwe23}.
The preliminary ALICE data, however, turn out to be a factor 1.72 higher than this prediction. {The reason for this discrepancies is presently not know to us, but
 should be investigated. We} now renormalize the distribution functions to the actual {preliminary mid-rapidity data in the eight centrality classes}.
 
 We determine the particle-production amplitude in all other centrality classes by assuming the same decreasing behavior as observed in $p$–Pb collisions at $\sqrt{s_{NN}} = 8.16\,$TeV, with parameter values given in table\ref{tab2}, and the centrality dependence of the mid-rapidity values for both $p$-Pb and $p$-O shown in Fig.\,\ref{fig2} in old and new centrality classes. 
\section{Results}
\label{sec:results}
We compare the calculated charged-hadron distributions in eight centrality classes to the preliminary 9.618 TeV $p$-O ALICE data \cite{uri25} in Fig.\,\ref{fig3}.
Positive pseudorapidities $\eta>0$ correspond to the forward ($p$-going) direction, while negative pseudorapidities indicate the backward (O-going) direction. 
Solid {blue} curves are the results from all three sources, with the absolute normalization taken from the {preliminary} ALICE data.
The dashed 
{orange} curves show charged hadrons (pions, kaons, protons, and their antiparticles) from the central gluon-gluon source, the dotted {green} curves are the ones from the respective $p$-going (forward) and O-going (backward) fragmentation sources. 
The dash-dotted {blue} curves are the distributions from our previous
prediction \cite{ksw25} -- albeit in the new centrality classes -- with the normalization taken from Monte Carlo calculations \cite{piwe23}. 

In our $p$-O model calculations, the amplitude of the charged-hadron production in the $p$-going direction becomes larger than the amplitude in the O-going direction in peripheral collisions. The central gluon-gluon source cannot compensate for this effect, although it shows a slight preference towards the O-going side even in peripheral collisions.
The {qualitative} effect is similar to our previous findings for $p$-Pb at 5.02 TeV \cite{sgw24}, where it matches actual ALICE data. To confirm the effect for $p$-O, data at larger values of pseudorapidity -- preferentially up to $|\eta|<5$ as in $p$-Pb -- will be required. 

{It is emphasized that the ALICE preliminary data that we compare with are not considered final results. Hence, we avoid extracting very precise quantitative conclusions that could become invalid after the final publication. As an example, we do not highlight a discrepancy between model and data that appears for central collisions in the positive pseudorapidity region}.
\section{Conclusions}
In this Letter, we have compared our previous predictions for centrality-dependent charged-hadron production in $p$-O collision at $\sqrt{s_{NN}}=9.618 \,$TeV with preliminary mid-rapidity ALICE data. The parameters of our three-source relativistic diffusion model with color-glass initial conditions have been adapted to these data, such that a prediction of the charged-hadron pseudorapidity distribution functions at larger rapidity values becomes feasible. We have calculated the full distribution functions in eight centrality classes, in accordance with the ALICE measurement. 

As in our previous model calculations for $p$-Pb, we find in $p$-O an inversion of the maximum particle-production amplitude from the O-going side in central collisions, to the $p$-going side in peripheral collisions. We interpret this inversion as being due to the strong gluon field in the O-going fragment that interacts with the valence quarks of the forward-going proton,
compared to the interaction of the valence quarks in the oxygen overlap zone with the much  weaker gluon field in the proton.
It will be interesting to see whether this prediction is confirmed once the final $p$-O data become available, which will then also allow for an
optimization to determine the model parameters more precisely.

\textit{Note added in proof}:
Recently published ATLAS min. bias 9.62 TeV $p$-O data (\textit{Phys. Rev. Lett.,} \textbf{137} (2026) 121901) have a lower absolute normalization than the preliminary ALICE data, but no centrality dependence. 
\acknowledgments
CL is grateful to Philipp Schulz for adapting his $p$-Pb  \texttt{C++} code to $p$-O, and for updating the centrality bins. 
\textit{Data availability statement}: The preliminary ALICE data that support our findings are available in Ref. \cite{uri25}.
\bibliographystyle{eplbib}

\bibliography{gw_26.bib}


\end{document}